\documentclass[cits]{PoS}
\usepackage{graphicx}

\title{The contribution of INTEGRAL to blazar science}

\ShortTitle{Blazars with INTEGRAL}

\author{\speaker{Luigi Foschini}\thanks{A great thank you for comments, notes and suggestions to (in alphabetical order): V. Beckmann, E. Bottacini, M. Chernyakova, M. Chiaberge, W. Collmar, L. Costamante, T.J.-L. Courvoisier, M. Giroletti, M. Gliozzi, G. Lichti, L. Ostorero, E. Pian, R.M. Sambruna, M. T\"urler, M. Villata.} \, and Valentina Bianchin\\
        INAF/IASF-Bologna, Via Gobetti 101, 40129, Bologna (Italy)\\
        E-mail: \email{foschini@iasfbo.inaf.it}}

\abstract{We review 6 years of \emph{INTEGRAL} observations of blazars, from Target-Of-Opportunity (TOO) to normal observations to coordinated campaigns, from the new and unexpected discoveries to the improvements in this research field. We also shortly review what is expected from \emph{INTEGRAL} for the forthcoming years.}

\FullConference{7th INTEGRAL Workshop\\
		 September 8-11 2008\\
		 Copenhagen, Denmark}

\begin{document}

\section{Introduction}
It is generally accepted that blazars are active galactic nuclei with a relativistic jet pointed toward the Earth with small angles ($<10^{\circ}$), so that the relativistic motion can account for negligible $\gamma-$ray attenuation \cite{URRYPADO}. Blazars are organized along a ``sequence'' depending on the characteristics of their spectral energy distribution (SED), where flat-spectrum radio quasars (FSRQ) lie on the most luminous side and BL Lac Objects define the low luminosity side \cite{FOSSATI,GHISELLINI1}. Several challenges were posed to the sequence (e.g. \cite{PADOVANI}) leading to a recent general revision and improvement \cite{GHISELLINI2,MARASCHI}. Specifically, while the original sequence was based on observational inputs (bolometric luminosity, shape of the SED), the recent improvements found their pillars in the physical parameters, like the mass of the central spacetime singularity and its accretion rate \cite{GHISELLINI2}. 

\begin{figure}[!t]
\centering
	\includegraphics[angle=270,scale=0.5]{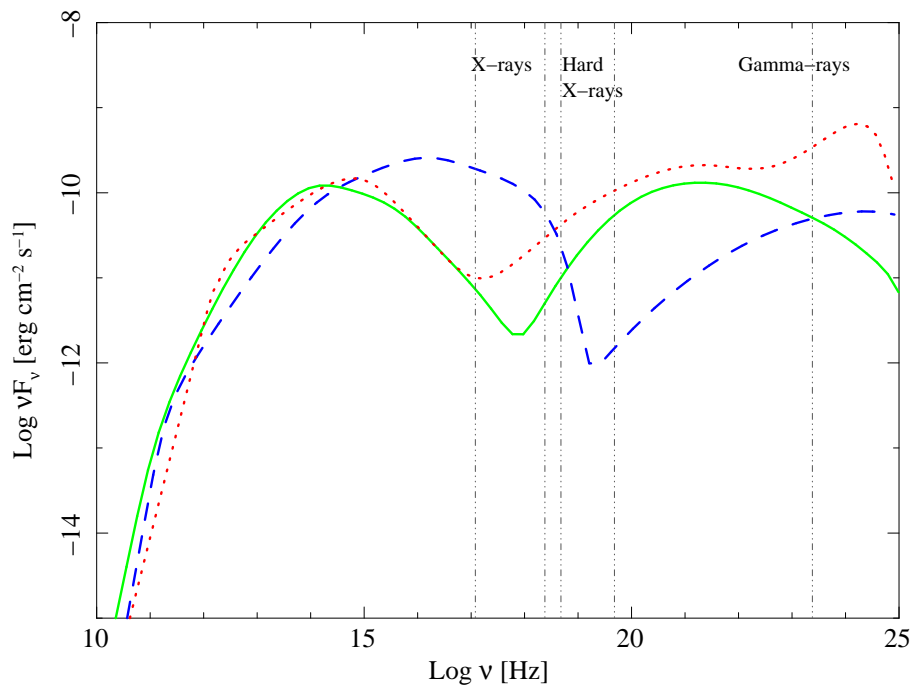}
	\caption{Typical SED of blazars: the red dotted line is the typical SED of Flat-Spectrum Radio Quasars (FSRQ); the green continuous line is for Low-Frequency Peaked BL Lac Objects (LBL); the dashed blue line is for High-Frequency Peaked BL Lac Objects (HBL). Grey vertical lines separate the X-rays ($E\approx 0.5-10$~keV: e.g. \emph{XMM-Newton}, \emph{Swift}, \emph{Chandra}, \emph{Suzaku}), hard X-rays ($E\approx 20-200$~keV: e.g. \emph{INTEGRAL}, \emph{Swift}, \emph{Suzaku}) and $\gamma-$rays domains (MeV-GeV-TeV ranges: \emph{Fermi}, \emph{AGILE}, \emph{MAGIC}, \emph{HESS}, \emph{VERITAS}, \emph{CANGAROO}). Example models are courtesy of G.~Ghisellini.}
	\label{fig:SED}
\end{figure}

However, despite of this general view, there are still several open questions to understand the nature of these sources, like:

\begin{itemize}
	\item where and how $\gamma-$rays are generated;
	\item what is the dependence of the radiation emission on the viewing angle (and therefore what is the link with radiogalaxies, if any);
	\item what is the jet composition;
	\item how the jet is coupled with the accretion disk;	
	\item what are the scaling laws (link with microquasars);
\end{itemize}

\noindent just to mention a few. Observations in the X-rays and hard X-rays are crucial to investigate these issues, since they probe the part of the SED where synchrotron and inverse-Compton emissions are competing (see Fig.~\ref{fig:SED}). 

Here we review the main contributions of \emph{INTEGRAL} \cite{INTEGRAL} to the blazar science, obtained during the observing campaigns performed from the launch on October $17$, $2002$, to date.

\section{The INTEGRAL catalogs of point sources}
Two point source catalogs built with \emph{INTEGRAL} data are currently available \cite{KRIVONOS,BIRD}. The first one contains $400$ sources and it is a true all sky survey, since it includes all the data from December $2002$ (rev $25$) to June $2006$ (rev $463$) plus a series of $13$ exposures explicitly required during AO periods to complete the all-sky survey ($200$~ks, PI Churazov) \cite{KRIVONOS}. The second catalog contains $421$ sources detected in the period from November $2002$ (rev $12$) to April $2006$ (rev $429$) \cite{BIRD}. We integrated these lists by searching in the available literature, since several blazars have been recently observed with specific campaigns and therefore are not yet in these catalogs. In addition, there are also a few cases of blazars not present in the catalogs and, probably, the reason is that since they are generally built by performing source detection on very long exposure maps, this resulted in missing the short-term variability related to strong flares (if the blazar has too low flux when in quiescence and exceeds the instrument detection threshold only during short timescales). 

A summary of all the blazars detected to date with \emph{INTEGRAL} is reported in Table~\ref{tab:blazars}.

\section{Some specific notes}
Here we shortly review the studies of specific or well-known sources based on \emph{INTEGRAL} data. These notes do not match exactly the list in Table~\ref{tab:blazars}, since not all the blazars detected by \emph{INTEGRAL} resulted in one or more dedicated papers. The source list is reviewed in order of ascending right ascension.

\subsection{PKS 0537-286 (z=3.1)}
This is the second highest redshift FSRQ detected by \emph{INTEGRAL} (the farthest is MG3~J$225155+2217$, see Sect.~$3.15$). Preliminary results presented at the present workshop showed a weak detection ($5\sigma$) in the hard X-ray band and a spectral flattening at soft X-ray energies \cite{BOTTACINI2}.

\begin{table}[t]
\centering
\begin{tabular}{llccl}
	\hline
	\textbf{Blazar}        & \textbf{Type}    & \textbf{Krivonos' Catalog} & \textbf{Bird's Catalog} & \textbf{Other} \\
  {}            & {}      & \cite{KRIVONOS}   & \cite{BIRD}	   & \textbf{References}\\
 	\hline
	RGB J0035+598 &  HBL   &  Yes         & Yes            & \\
	PKS 0537-286  &  FSRQ  &   No          & No             & \cite{BOTTACINI2}\\
	S5 0716+714   &  LBL   &  No          & No             & \cite{FOSCHINI1, OSTORERO, PIAN1}\\
	S5 0836+71    &  FSRQ  &   Yes         & Yes            & \cite{PIAN1}\\
	Mkn 421       &  HBL   &  No          & No             & \cite{LICHTI}\\
	4C 04.42      &  FSRQ  &   Yes         & Yes            & \cite{DEROSA2}\\
	3C 273        &  FSRQ  &   Yes         & Yes            & \cite{BIANCHIN,CHERNYAKOVA,COURVOISIER,TURLER,SOLDI}\\
	PKS 1241-399  &  FSRQ  &   Yes         & No             & \\	
	3C 279        &  FSRQ  &   Yes         & Yes            & \cite{BOTTCHER,COLLMAR1,COLLMAR2}\\	
	1ES 1426+428  &  HBL  &   No          & No             & \cite{WOLTER}\\
	Swift J1656.3-3302 & FSRQ   & No          & No             & \cite{MASETTI}\\
	Mkn 501       &  HBL  &   Yes         & No             & \\
	NRAO 530      &  FSRQ &    No          & No             & \cite{FOSCHINI2}\\
	PKS 1830-211  &  FSRQ &    Yes         & Yes            & \cite{DEROSA1,ZHANG}\\
	PKS 1921-293  &  FSRQ &    No          & Yes            & \\
	1ES 1959+650  &  HBL  &   No          & No             & \cite{BOTTACINI}\\
	PKS 2149-307  &  FSRQ &    No          & No             & \cite{BIANCHIN2}\\
	BL Lac        &  LBL  &   Yes         & Yes            & \\
	3C 454.3      &  FSRQ &    Yes         & No             & \cite{PIAN2,VERCELLONE}\\
	MG3 J225155+2217 &  FSRQ & Yes         & No             & \cite{BASSANI,MARASCHI}\\
	\hline
\end{tabular}
	\caption{Summary of blazars detected with \emph{INTEGRAL} during its $6$ years of activity.}
	\label{tab:blazars}
\end{table}

 \subsection{S5 0716+714 (z=0.31)}
 This LBL was the target of two multiwavelength campaigns. The first one, performed in November $2003$, was organized by the ENIGMA-WEBT network (PI S. Wagner), when the source was extremely bright at radio frequencies, but optically faint ($R=14.17-13.64$). No detection at hard X-rays was found \cite{OSTORERO}, but it is worth noting that just a few days before this campaign, \emph{INTEGRAL} was knocked down by a strong solar flare, which caused the instruments to switch off. 
 
 The second campaign was instead performed in April $2004$, when high optical flux close to the historical maximum ($R=12.1$) recorded at the end of March $2004$, triggered simultaneous observations with \emph{INTEGRAL} (PI Pian) and \emph{XMM-Newton} (PI Tagliaferri). In this case, a faint detection ($4.5\sigma$) was measured in the $30-60$~keV energy band \cite{PIAN1}, consistent with the extrapolation from \emph{XMM-Newton} data \cite{FOSCHINI1}. The studies on this outburst led to the conclusions that S5~$0716+714$ displayed two types of variability on long and short time scales \cite{FOSCHINI1}. On long time scales, there is only a change in flux without spectral variations. The gradual decay after the outburst can be due to the escape of electrons from the processing region or to a decrease of the seed photons. Modeling of the SED built with data obtained in $2004$ and comparison with previous observations in quiescence (1996, see \cite{GIOMMI99}) revealed that the physical parameters had negligible changes, except for the injected power. On the other hand, the short time scale variability is characterized by flux and spectral changes and it is likely to be due to variations in the slope of the electron distribution.
 
Interestingly, VLBI observations at $15$~GHz showed that, during the November $2003$ campaign (no hard X-rays detection), the blazar core had the highest flux, while, in April $2004$ (second campaign; detection with \emph{INTEGRAL}), there was a peak in the emission from the knot A \cite{BRITZEN}.

\subsection{S5 0836+710 (z=2.172)}
The observation of this high-redshift flat-spectrum radio quasar was a ``gift'' from the large field of view (FOV~$=29^{\circ}\times 29^{\circ}$ at zero response) of the IBIS imager onboard \emph{INTEGRAL}. Indeed, S5~$0836+710$ is $6.5^{\circ}$ far from S5~$0716+714$, which was the target of the two above mentioned campaigns, and therefore it was detected as well. Comparison with previous observations with \emph{BeppoSAX} \cite{TAVECCHIO}, indicate that the flux was about a factor 3 lower, but the photon index remained constant within the measurement errors \cite{PIAN1}.

\subsection{Mkn 421 (z=0.03)}
The \emph{INTEGRAL} campaign of June $2006$ (PI Lichti) on this famous HBL was triggered by an exceptional X-ray flux (up to $80$~mcrab measured by the All-Sky Monitor of \emph{RXTE} in the $2-10$~keV energy band). The source was significantly detected with all the instruments onboard \emph{INTEGRAL} (except SPI, which was annealing) and the SED, built with multiwavelength data, clearly showed the shift of the synchrotron peak from soft to hard X-rays \cite{LICHTI}. The study of the variability across about $10$ days of exposure revealed several flares, but only one showed significant spectral changes \cite{LICHTI}.

\subsection{4C 04.42 (z=0.965)}
The \emph{XMM-Newton} and \emph{INTEGRAL} observation of this FSRQ revealed a hard spectrum ($\Gamma \sim 1.2$) extending to high-energy and a soft excess at low energies ($E<2$~keV), which was interpreted as evidence of bulk motion Comptonization \cite{DEROSA2}.

 \subsection{3C 273 (z=0.158)}
 3C 273 is perhaps the most ever known blazar, since the historical observations by Marteen Schmidt in 1963 \cite{SCHMIDT} (see also \cite{SOLDI} for a recent review). It was the first blazar observed by \emph{INTEGRAL}, after the Performance Verification Phase on January 5, 2003 (rev 28, PI Courvoisier) \cite{COURVOISIER}. This source was periodically observed during the \emph{INTEGRAL} lifetime, often in coordination with \emph{XMM-Newton} and \emph{RXTE} \cite{CHERNYAKOVA}. The available data over the period $2003-2005$ suggest a possible anticorrelation between hard ($20-40$~keV) and soft ($0.2-1$~keV) X-rays \cite{BIANCHIN}. In $2004$, during a historical minimum at submm wavelengths, a broad iron emission line was detected at $6.4$~keV \cite{TURLER}. 
 
 \subsection{3C 279 (z=0.536)}
 This is another well known blazar and was the target of two multiwavelength campaigns: one in 2003 and the other in 2006 (PI Collmar in both cases) \cite{COLLMAR1,COLLMAR2}. While the first campaign was planned independently from the status of the source, the second one was triggered by high optical flux observed by ground telescope at the beginning of January $2006$ (see also \cite{BOTTCHER}). By comparing the SED obtained from the two campaigns, the most striking feature is that, despite a strong change in optical fluxes (about one order of magnitude), little or negligible changes occurred in hard X-rays \cite{COLLMAR2,BOTTCHER}. The theoretical implications are still to be understood. 

\begin{figure}[!t]
\centering
	\includegraphics[angle=270,scale=0.45]{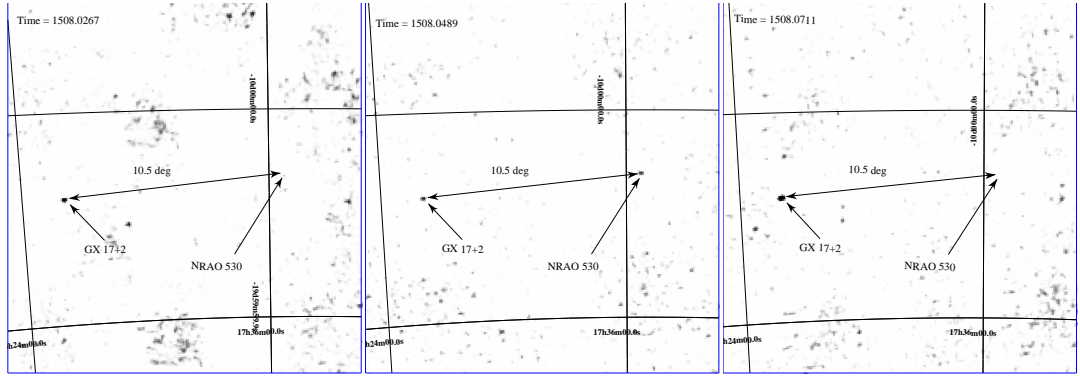}
	\caption{The short and strong hard X-ray flare observed from NRAO~$530$ on February $17$, $2004$. This is a sequence of three consecutive pointings ($\approx 2$~ks each) of IBIS/ISGRI in the $20-40$~keV energy band, showing that the source suddenly appeared and disappeared (from \cite{FOSCHINI2}).}
	\label{fig:NRAO530}
\end{figure}

\subsection{1ES 1426+428 (z=0.129)}
This is an extreme HBL, with the synchrotron emission probably peaking in the hard X-rays. \emph{INTEGRAL} observation performed in May~$2006$ (PI Wolter) showed a rather flat photon index for a HBL ($\Gamma \sim 2$), confirming its unusual characteristics \cite{WOLTER}.

\subsection{Swift J1656.3-3302 (z=2.40)}
This high-redshift blazar was discovered by \emph{Swift} and later confirmed by \emph{INTEGRAL}. The high-energy part of the spectrum displayed a photon index of about $1.6$, but with a spectral flattening at low energies ($E<2$~keV), which was interpreted as absorption intrinsic to the source \cite{MASETTI}.

\subsection{NRAO 530 (z=0.902)}
The importance of having a large FOV is again underlined in this case: NRAO~$530$ had a strong and short flare that was serendipitously detected by \emph{INTEGRAL} during the normal operations of the Galactic Centre Deep Exposure on February $17$, $2004$. The $20-40$~keV flux emitted by the blazar increased up to $\approx 2\times 10^{-10}$~erg~cm$^{-2}$~s$^{-1}$ and then dropped below the instrument detection limit in less than $1$ hour (see Fig.~\ref{fig:NRAO530}) \cite{FOSCHINI2}. Given such extraordinary behaviour, never observed in a flat-spectrum radio quasar, there was the doubt of a contaminating source within the $3'$ error circle of IBIS. However, a couple of snapshots with the X-Ray Telescope (XRT) onboard \emph{Swift} revealed that NRAO~$530$ is the only high-energy source present in the IBIS error circle. In addition, this blazar is known to display high-amplitude variability at almost all the wavelengths and therefore, the flux measured by \emph{INTEGRAL} is not at odds with historical observations.

\subsection{PKS 1830-211 (z=2.507)}
This is a gravitationally lensed FSRQ and therefore the jet radiation, in addition to the Doppler boosting, is amplified also by the gravitation of the intervening galaxy at $z=0.89$. The analysis of the SED built with \emph{Chandra}, \emph{INTEGRAL} and EGRET data (not simultaneous) favours the typical emission mechanism powering FSRQ, with a spectral flattening at soft X-rays ($E<4$~keV) that was initially interpreted as ionized absorption intrinsic to the source \cite{DEROSA1}. However, a more recent study on a SED built with \emph{XMM-Newton}, \emph{INTEGRAL}, COMPTEL and EGRET data (not simultaneous) with the highest statistics available to date, strongly supported the hypothesis that the soft X-ray spectral flattening is the natural curvature of the FSRQ continuum \cite{ZHANG}.

\subsection{1ES 1959+650 (z=0.047)}
This HBL was detected during the observations of the Key Programme on the North Ecliptic Pole region, with a flux about one half of that displayed in the $2001$ outburst, but with a hard photon index of $\approx 1.9$ \cite{BOTTACINI}.

\begin{figure}[!t]
\centering
	\includegraphics[scale=0.45]{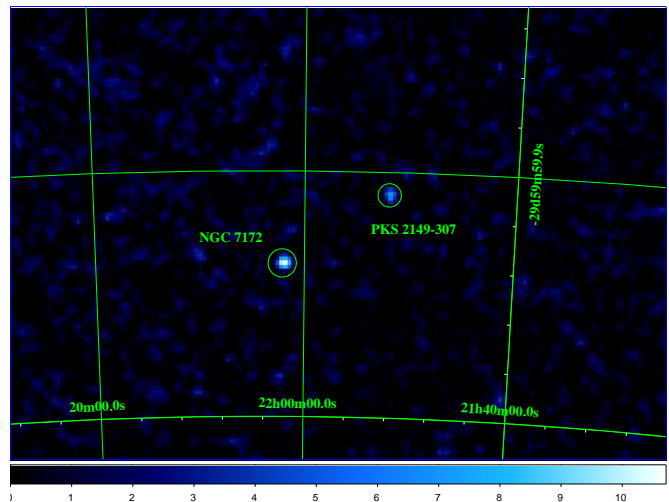}
	\caption{IBIS/ISGRI significance map in the $20-40$~keV energy band of the region around NGC~$7172$ with the serendipitous detection of PKS~$2149-307$.}
	\label{fig:PKS2149}
\end{figure}

\subsection{PKS 2149-307 (z=2.345)}
This high-redshift blazar was detected serendipitously in $2004$ during an observation of the nearby Seyfert~$2$ active nuclei NGC~$7172$ (distance $\approx 2.6^{\circ}$, see Fig.~\ref{fig:PKS2149}). The \emph{INTEGRAL}/IBIS spectrum in the $20-100$~keV has the same shape of that detected in $2005$ by \emph{Swift}/BAT during its first nine month of operations \cite{SAMBRUNA2}, but it is a factor two lower \cite{BIANCHIN2}. Therefore, over about one year, the hard X-ray flux of this blazar roughly doubled its value, without changes in slope.

\subsection{3C 454.3 (z=0.859)}
This FSRQ has been extremely active since $2005$ and several multiwavelength campaigns were performed. In May $2005$, 3C~$454.3$ underwent a long and intense outburst observed by many ground based telescopes (WEBT \cite{VILLATA}) and satellites (\emph{Swift} \cite{GIOMMI}, \emph{Chandra} \cite{VILLATA}). Also \emph{INTEGRAL} was activated and pointed to the blazar on May~$15-18$, $2005$ (PI Pian) \cite{PIAN2}. This observation can be better understood in the framework of the ``jet economic model'' \cite{ECONOMODEL} and by comparing the source evolution over several years \cite{GHISELLINI3}. According to this model, the ratio between Synchrotron Self-Compton (SSC) and External Compton (EC) emission is regulated by the bulk Lorentz factor $\Gamma$, which in turn is linked to the compactness of the source. This means that changes in the shape of the SED can be explained with an almost constant jet power and with different places where the main dissipation occurs. In $2005$, the high hard X-rays flux observed by \emph{INTEGRAL} can be modeled with low $\Gamma$ and high magnetic field, favouring the hypothesis that the main dissipation occurred in a compact region, where the SSC dominates and the EC gave little contribution. Instead, in $2007$, a high $\gamma-$ray flux observed by \emph{AGILE} (\cite{BULGARELLI, VERCELLONE1}) together with moderately low optical and X-ray fluxes, suggest that the dissipation occurred at larger distances, therefore with high $\Gamma$ and low magnetic field \cite{GHISELLINI3}. 

3C~$454.3$ continued to be active in $2007$ (and it is still very active also in $2008$, displaying an exceptional outburst observed by \emph{Fermi}/LAT \cite{TOSTI}), so that another multiwavelength campaign involving \emph{INTEGRAL} was triggered by the \emph{AGILE} team in November $2007$ (PI Vercellone). The $20-200$~keV flux was about one order of magnitude less than that detected in $2005$, while the $\gamma-$ray flux was quite high \cite{VERCELLONE}, favouring again the ``jet economic model'' interpretation.

\subsection{MG3~J225155+2217 (z=3.668)}
\begin{figure}[!t]
\centering
	\includegraphics[scale=0.45]{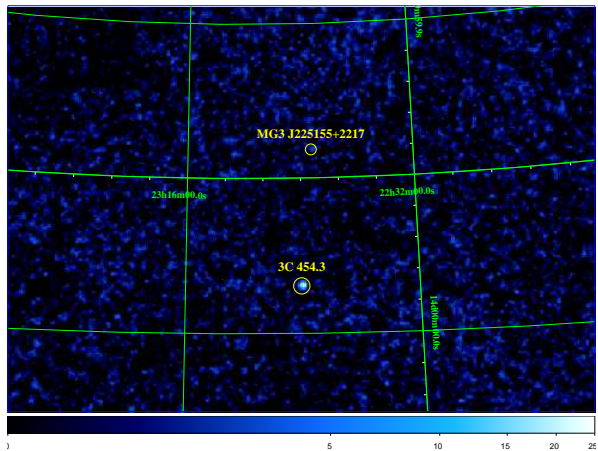}
	\caption{IBIS/ISGRI significance map in the $20-40$~keV energy band of the region around 3C~$454.3$ and with the blazar MG3~J$225155+2217$.}
	\label{fig:3CMG3}
\end{figure}

Yet another gift from the large FOV of \emph{INTEGRAL}. This source was first serendipitously detected in the FOV of 3C~$454.3$ observation in May $2005$ (distance $\approx 6^{\circ}$; see Fig.~\ref{fig:3CMG3}) and reported as unidentified source in the Krivonos' Catalog \cite{KRIVONOS}. Later, it was associated with the high-redshift ($z=3.668$) blazar MG3~J225155+2217 \cite{BASSANI} and it is the farthest detected by \emph{INTEGRAL} to date. The modeling of the SED showed that the emission from this source is largely dominated by the inverse-Comtpon component \cite{MARASCHI}, but still in the framework of a typical high-redshift FSRQ (like, e.g., Swift~J$0746.3+2548$ at $z=2.979$ \cite{SAMBRUNA}).

\section{Final remarks}
During about $6$ years of activity, \emph{INTEGRAL} detected $20$ blazars, with a strong preference to FSRQ (13 sources) and, particularly, $6$ FSRQ have redshift greater than $2$. The major strength of \emph{INTEGRAL} is the large FOV of its instruments (specifically IBIS, with more than $800$ square degrees), which allowed us to perform some important serendipitous discoveries, like the high-redshift extreme blazar MG3~J$225155+2217$ and the short and strong flare of NRAO~$530$. The short term variability can be studied only in a few cases, for the brightest sources, but the sensitivity that can be reached with long/deep exposures is sufficient to probe the moderately high-redshift Universe. These assets can be emphasized by the Key Programmes, which are now (from AO7) constituting the $70\%$ of the observing time.

\emph{INTEGRAL} gave interesting contributions also in studying blazars in outburst, although its contribution is ``diluted'' in the multiwavelength campaign. Nevertheless, great synergies are expected with \emph{Fermi}, \emph{AGILE}, \emph{MAGIC II} and the forthcoming \emph{HESS II}. 

For the studies of hard X-ray emission from blazars beyond $2012$, the year when \emph{INTEGRAL} is planned to be switched off, it is expected that its legacy will be received by \emph{Simbol-X} and, hopefully, by the \emph{International X-ray Observatory} (IXO).

\end{document}